\documentclass[doublecol]{epl2}

\newcommand{\be}{\begin{eqnarray*}}
\newcommand{\ee}{\end{eqnarray*}}
\newcommand{\gl}[1]{(\ref{#1})}
\newcommand{\ta}[2]{ \frac{ {\mathrm{d}} #1 } {{\mathrm{d}} #2}}

\newcommand{\bee}{\begin{eqnarray}}
\newcommand{\eee}{\end{eqnarray}}

\title{NLO-QCD corrections to $W\gamma j$
production}

\author{
F.~Campanario\inst{1,2}
\and
C.~Englert\inst{1}
\and
M.~Spannowsky\inst{1}
\and 
D.~Zeppenfeld\inst{1}
}

\institute{   
\inst{1} Institute for Theoretical Physics, University of Karlsruhe, KIT, 76128 Karlsruhe, Germany\\
\inst{2} Departament de F\'isica Te\`orica and IFIC, Universitat de Val\`encia - CSIC, E-46100 
Burjassot, Val\`encia, Spain
}
 
\pacs{12.38.Bx} {Perturbative calculations}
\pacs{13.85.-t}{Hadron-induced high- and super-high-energy interactions (energy $>$ 10 GeV)}
\pacs{14.70.Bh}{Photons}
\pacs{14.70.Fm}{W bosons}
 
\abstract{
We calculate the $W^\pm\gamma j + X$-production cross sections at next-to-leading order QCD for Tevatron and LHC collisions. We include leptonic decays of the $W$ to light leptons, with all off-shell effects taken into account. The corrections are sizable and have significant impact on the differential distributions.}

\begin{document}

\maketitle

%
%
%
\section{Introduction}
At hadron colliders such as the CERN Large Hadron Collider (LHC) and the Fermilab Tevatron,
electroweak boson production in association with jets represents important signal processes
as well as backgrounds to future searches beyond the Standard Model (BSM). One example is the measurement of anomalous tri-boson couplings, arising from BSM physics, which can be obscured by higher-order QCD effects. For these searches, significance-improving strategies include jet-vetos, which amount to subtraction of a leading order cross section 
\cite{Baur:1993ir}, and are plagued at present with typical QCD-scale uncertainties. 
Improved QCD-precision of production cross sections is therefore essential and has been
agreed on as a common goal of precision phenomenology in the so-called ``Les Houches wish-list'' \cite{Buttar:2006zd}. Considerable
progress in completing this task has been accomplished, {\it cf.} \cite{Dittmaier:2007th,
Campanario:2008yg, Binoth:2008kt,Bredenstein:2009aj,Bevilacqua:2009zn}. Concerning electroweak boson production in association with a jet, the QCD-corrections to $W^+W^-+\hbox{jet}$ have been recently provided in \cite{Dittmaier:2007th}.

In this letter we examine $W^\pm\gamma+\hbox{jet}$ production at next-to-leading order (NLO) QCD, including leptonic decays of the $W^\pm$.
We devote special care to the development of a fully-flexible, numerically stable parton-level Monte Carlo
implementation, based on the \textsc{Vbfnlo}-framework \cite{Arnold:2008rz}. 
Although full leptonic decays of the massive $W^\pm$ are included, we will refer to the processes as 
$W^\pm\gamma j$ production in the following.

\begin{figure}
\onefigure[scale=0.4]{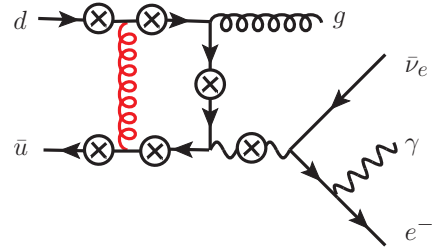}
\caption{Representative Feynman graph contributing to the virtual corrections to
the partonic subprocess $\bar u d\rightarrow  e^-\bar\nu_e \gamma g$ at
${\cal{O}}(\alpha^3\alpha_s^2)$. 
The crosses mark other points where the photon is attached to the quark line and
the $W$ boson.}
\label{fig:graphvirt}
\end{figure}

\section{Elements of the Calculation and Checks}
The leading order contribution, at ${\cal{O}}(\alpha^3\alpha_s)$, to the process
$pp\rightarrow \ell^-\bar\nu_\ell\gamma j+X$ includes
subprocesses of the type $q \bar Q\rightarrow \ell^-\bar \nu_\ell \gamma g$, and
$qg$ and $\bar Qg$ initiated subprocesses which are related by crossing.

The 10 Feynman graphs of each subprocess can be classified into
two categories: First, configurations where the photon is emitted from the 
$W$ or the $W$'s decay lepton, and, second, graphs where the photon is
emitted from the quark line. Performing the virtual correction at 
${\cal{O}}(\alpha^3\alpha_s^2)$, these topologies give rise to self-energy, triangle, box, 
and pentagon (sub-)diagrams. The loop corrections are treated
using standard methods: Self-energy, triangle, box and pentagon integrals 
are evaluated in terms of tensor coefficients \cite{Passarino:1978jh,Denner:2002ii} in dimensional reduction, after having applied $\overline\mathrm{MS}$-renormalization. 
We combine the virtual corrections to groups that include all loop diagrams derived from
a Born level configuration, i.e. all self-energy, triangle, box and pentagon corrections to a quark line with three attached gauge bosons are combined to a single routine. This method leaves us with a universal set of virtual building blocks, which are then assembled for the specific process under consideration. This strategy has already been applied to various phenomenological studies at NLO-QCD precision, {\it e.g.} \cite{Campanario:2008yg,Englert:2008wp}.
 
\begin{figure}
\onefigure[scale=0.4]{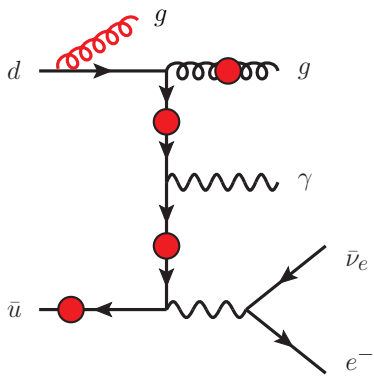}
\caption{Sample Feynman graph contributing to the partonic real emission subprocess
$\bar ud\rightarrow e^- \bar\nu_e \gamma g g$ at ${\cal{O}}(\alpha^3\alpha_s^2)$. 
The gluon is attached to the
quark and gluon lines at positions marked by the circles. Feynman graph topologies, where the photon is radiated off at different positions analogous to fig.~\ref{fig:graphvirt}, are not shown.}
\label{fig:graphrem}
\end{figure} 

The reduction of the loop diagrams has been calculated in two independent ways 
for verification reasons. The first approach uses in-house routines within in the framework of \textsc{FeynCalc} \cite{Mertig:1990an} and {\sc FeynArts}, 
while the second one relies on \textsc{FeynArts}, \textsc{FormCalc}, and \textsc{LoopTools}
\cite{Hahn:1998yk,Hahn:2000kx}, with modifications, in particular 
to the treatment of divergencies, as described in \cite{Bredenstein:2008zb}.
We find that 
both calculations numerically agree within \textsc{Fortran} precision for different 
points. Performing the NLO-computation in the chiral limit, the arising infrared (IR)
singularities have been determined separately in independent
approaches, and checked against existing results in the literature \cite{Dittmaier:2003bc,Figy:2007kv}.

The IR singularities encountered in the real emission contributions are regularized
using the Catani-Seymour dipole formalism \cite{Catani:1996vz}. 
The numerical 
implementation of the dipoles has been
numerically checked against \textsc{MadDipoles} \cite{Frederix:2008hu}.
The code is optimized such that intermediate dipole-results are stored and reused in order to 
avoid redundant calculations.
Remaining finite collinear
terms, after renormalizing the parton distribution functions according to
\cite{Catani:1996vz}, were analytically calculated in two independent ways. 
We integrate the finite collinear terms over the real emission phase space by appropriately 
mapping the LO-phase space, as done in \cite{Figy:2003nv}.
The cancellation of virtual IR singularities against the one-parton phase space--integrated dipoles'
has been checked analytically.

We evaluate the leading-order matrix element, as well as the subtraction terms using partly 
modified \textsc{Helas}-routines \cite{Murayama:1992gi} generated with {\textsc{MadGraph}} 
\cite{Alwall:2007st}. Due to the increase of subprocesses when going
to the evaluation of the IR-subtracted real emission matrix element, optimization is imperative in 
order not to jeopardize CPU time. Here, the matrix element is calculated using the
spinor helicity formalism of \cite{Hagiwara:1988pp}, 
and intermediate numerical results, common to all subprocesses, are stored and reused, thus speeding
up the numerical code.
The real emission matrix elements, {\it cf.} fig.~\ref{fig:graphrem} for sample graphs of the partonic subprocess
$\bar u d \rightarrow e^- \bar\nu_e \gamma g g$, have been checked 
numerically against code generated by \textsc{MadGraph} for every subprocess. Integrated 
results were checked against {\sc Sherpa}  \cite{Gleisberg:2008ta}. Tab.~\ref{tab:madevent} 
representatively gives the result of our comparison of integrated cross sections with 
\textsc{MadEvent} v4.2.21 and \textsc{Sherpa} v.1.1.3 for the leading order and the real 
emission dijet contribution, {\it i.e.} the process $p p\rightarrow e^-\bar{\nu}_e \gamma jj+X$ 
at ${\cal{O}}(\alpha^3\alpha_s^2)$, for cuts specified below.

\begin{table}[t]
\begin{center}
\begin{tabular}{r c c }
\hline
&  $W^-\gamma j$ [fb]&  $W^-\gamma jj$ [fb]\\
\hline
mod. \textsc{Vbfnlo} & $268.38 \pm 0.12$ & $124.74 \pm 0.10$ \\
\textsc{Sherpa}   &     $268.14\pm 0.37$ & $124.35\pm 0.59$ \\
\textsc{MadEvent}  &  $268.24\pm 0.69$ & $123.80\pm 0.40$ \\
\hline
\end{tabular}
\caption{Comparison of integrated $W^-\gamma j$ and $W^-\gamma jj$ tree-level cross sections at the LHC. The cross sections were calculated with our modified version of \textsc{Vbfnlo}, \textsc{MadEvent} v4.4.21, and \textsc{Sherpa} v.1.1.3. The QCD-IR-safe photon-isolation is replaced by a conventional separation \mbox{$R_{j\gamma}\geq 1$} for all jets. We also require $R_{\ell\gamma} \geq 0.4$ and $R_{jj}\geq 0.7$. All other parameters and cuts are chosen as described in the text.}
\label{tab:madevent}
\end{center}
\end{table}

Concerning the Monte-Carlo implementation of the virtual corrections, we have implemented
the loop contributions using our \textsc{Vbfnlo} routines, that 
involve the Passarino-Veltman reduction scheme \cite{Passarino:1978jh} up to boxes, the Denner-Dittmaier reduction scheme \cite{Denner:2002ii} for pentagons, and the spinor helicity formalism of \cite{Hagiwara:1988pp}. Throughout, the numerical integration is performed using a modified version of \textsc{Vegas} \cite{Lepage:1977sw}, which is part of the \textsc{Vbfnlo} package, with different channels for the two- and three-body decay of the $W$ boson. 
Finite width effects of the $W$ boson are taken into account
using a modified version of the complex mass scheme of
\cite{Denner:1999gp}: The weak mixing angle is taken to be
real, while using a Breit-Wigner-propagator for the $W$ boson.
This scheme corresponds to the implementation in \textsc{MadGraph}.

For a more detailed discussion of the calculation and its numerical implementation, we
refer the reader to a separate paper \cite{toapp}.
 
\section{Numerical Results}
We use CTEQ6M parton distributions \cite{Pumplin:2002vw} with $\alpha_s(m_Z)=0.118$ at NLO, and the CTEQ6L1 set at LO. We choose $m_{Z}=91.188~\rm{GeV}$, $m_{W}=80.419~\rm{GeV}$ and $G_F=1.16639\times 10^{-5}~\textnormal{GeV}^{-2}$ as electroweak input parameters and derive the electromagnetic coupling $\alpha$ and the weak mixing angle $\sin\theta_w$ via Standard Model-tree level relations. The center-of-mass energy is fixed to $14~\rm{TeV}$ for LHC and $1.96~\rm{TeV}$ for Tevatron collisions, respectively. We only consider $W^\pm$ decays to one family of light leptons, {\it e.g}. $W^-\rightarrow e^-\nu_e$, and treat these leptons as massless.
The CKM-matrix is taken to be diagonal, and we neglect bottom contributions throughout. A non-diagonal CKM matrix decreases our leading-order LHC-result at the per mill level as gluon-induced processes dominate the cross section. The correction for the Tevatron results, which are mostly
quark-induced, is about 3\%. 
These corrections are well below the residual scale dependence at NLO-QCD. The bottom contributions are negligible and can be further suppressed by b-tagging.
Jets are recombined via the $k_T$ algorithm \cite{Catani:1993hr} from massless partons of pseudorapidities $|\eta|\leq 5$ with resolution parameter $D=0.7$. The jets are required to lie in the rapidity range $|y_j|\leq 4.5$ with 
$p_T^{\mathrm{jet}}\geq 50~\rm{GeV}$. The photon and the charged lepton are chosen to be rather hard and central, $p_{T}^\ell \geq 20 ~\rm{GeV}$, $p_{T}^\gamma \geq 50 ~\rm{GeV}$, 
$|\eta_\ell|,|\eta_\gamma|\leq
2.5$, while being separated in the azimuthal angle-pseudorapidity plane by 
\mbox{$R_{\ell\gamma} = (\Delta \phi_{\ell\gamma}^2 + \Delta\eta_{\ell\gamma}^2)^{1/2}\geq 0.2$}. 
For the separation of the charged lepton from observable jets, we choose $R_{\ell j}\geq 0.2$. A naive isolation criterion for the partons and the photon spoils IR-safety, yet isolation is necessary to avoid fragmentation contributions. We apply the method suggested in \cite{Frixione:1998jh}, demanding
\bee
\label{photonisolation}
\sum_{i, R_{i\gamma}<R} p_T^{{\rm{parton}}, i} \leq
\frac{1- \cos R}{1- \cos \delta_0}\,  p_T^\gamma \qquad \forall R\leq \delta_0,
\eee
\noindent where the index $i$ runs over all partons, found in a cone around the photon of size $R$. 
For the cut-off parameter, that determines the QCD-IR-safe cone size around the photon, we choose $\delta_0=1$. 

\begin{figure}[t]
\onefigure[scale=0.85]{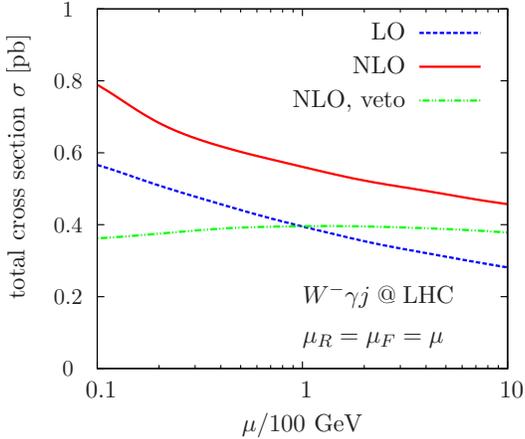}
\caption{Comparison of the scale dependence of the total cross section of $pp\rightarrow
e^-\bar\nu_e \gamma j+X$ at LO (dashed), NLO-QCD (solid), and NLO-QCD with the second jet vetoed (dot-dashed)
for the cuts chosen as described in the text at the LHC.}
\label{fig:crossecs}
\end{figure}
\begin{figure}
\onefigure[scale=0.85]{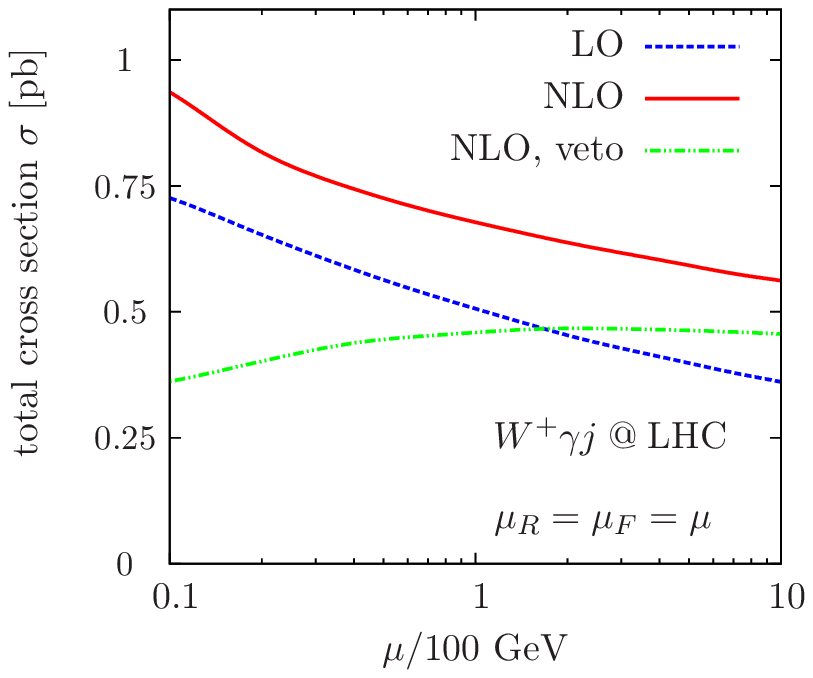}
\caption{Comparison of the scale dependence of the total cross section of $pp\rightarrow
e^+\nu_e \gamma j+X$ at LO (dashed), NLO-QCD (solid), and NLO-QCD with the second jet vetoed (dot-dashed) for the cuts chosen as described in the text at the LHC.}
\label{fig:crossecswp}
\vspace{1.05cm}
\onefigure[scale=0.85]{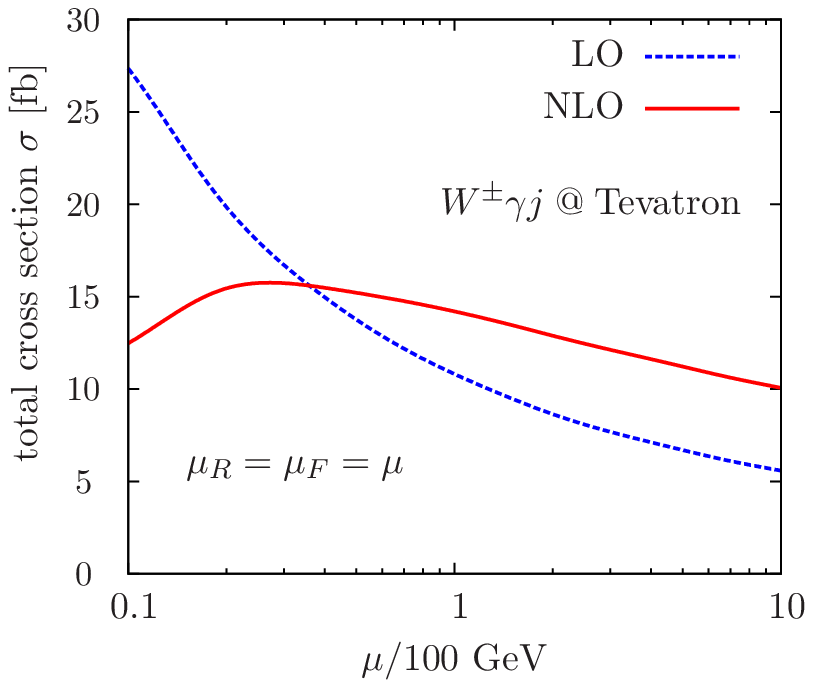}
\caption{Tevatron-comparison of the scale dependence of the total cross section of $p\bar p\rightarrow
e^-\bar\nu_e \gamma j+X$ or $p\bar p\rightarrow e^+\nu_e \gamma j+X$ at LO (dashed) and 
NLO-QCD (solid)
for the cuts chosen as described in the text.}
\label{fig:crossecstev}
\end{figure}

\begin{figure}[h!]
\onefigure[scale=0.85]{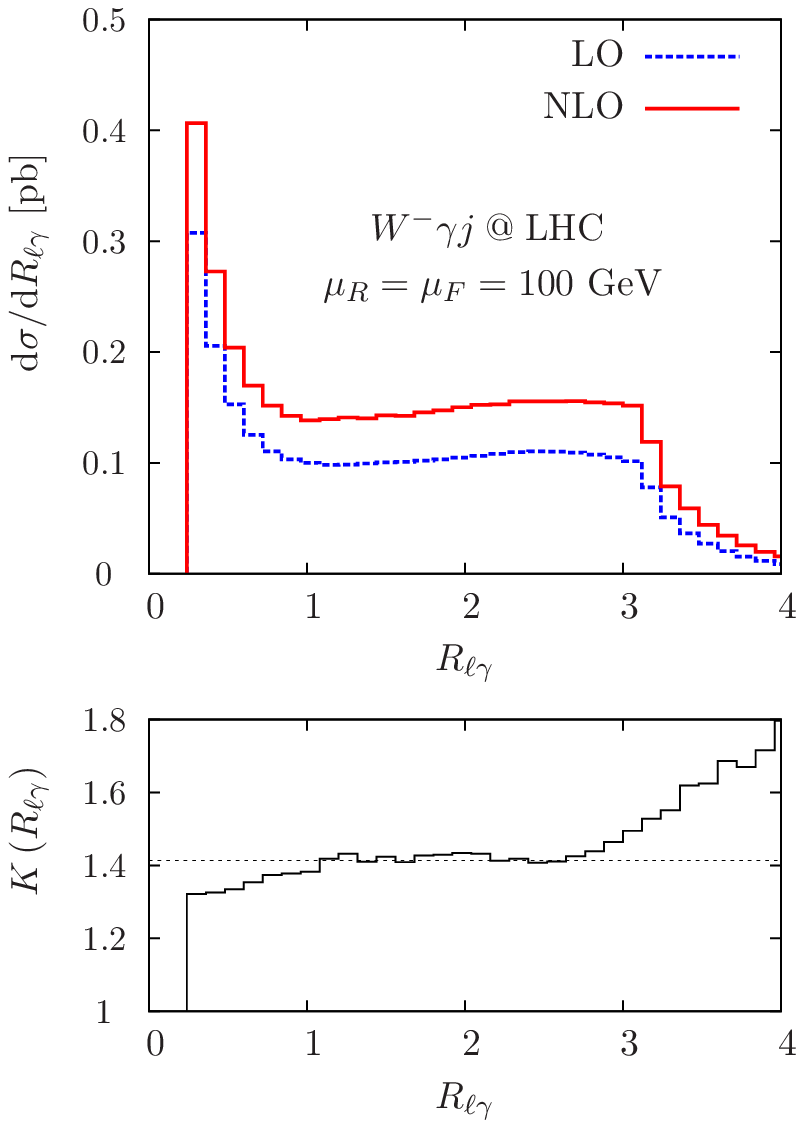}
\caption{Differential distribution of the photon-lepton separation $R_{\ell \gamma}$ at LO (dashed) and at NLO (solid). The lower panel shows
the differential $K$-factor. The dotted line denotes the total $K$-factor of Tab. \ref{tab:kfactors}.
}
\label{fig:ral}
\vspace{0.57cm}
\onefigure[scale=0.85]{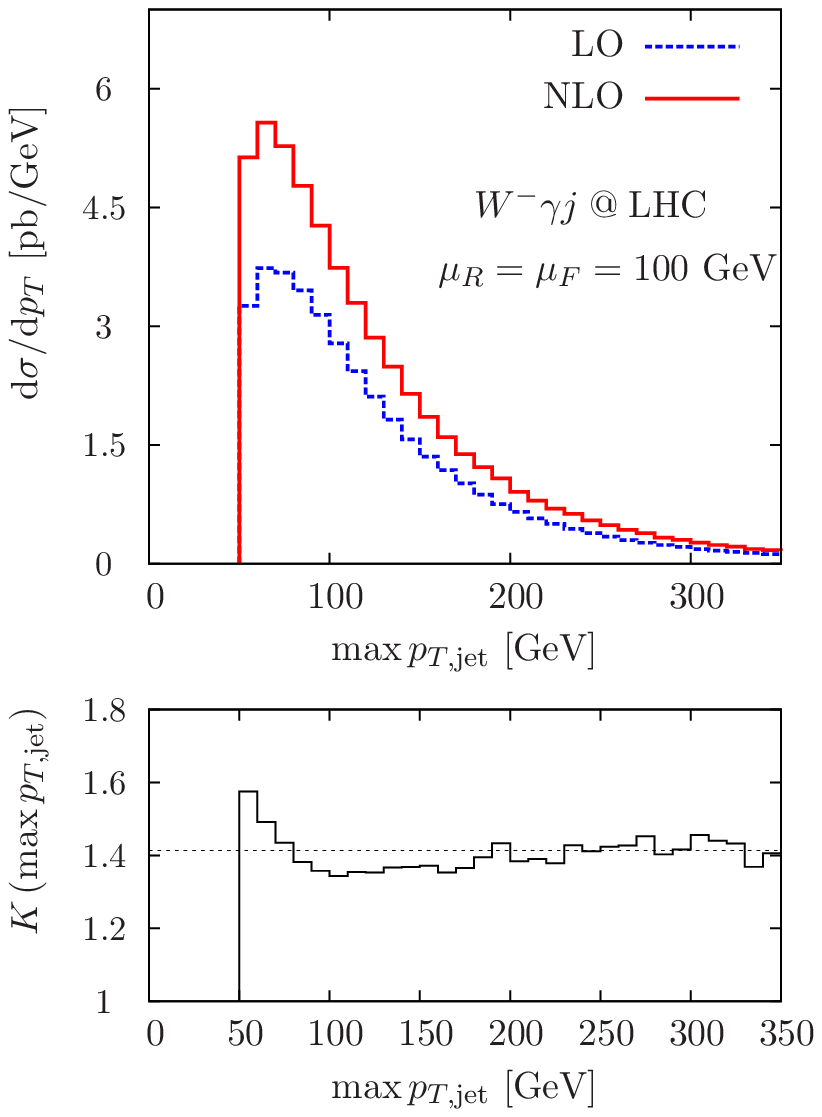}
\caption{Maximum jet-$p_T$ distribution at leading order (dashed) and next-to-leading order QCD (solid). The dotted line denotes the total $K$-factor of Tab. \ref{tab:kfactors}.}
\label{fig:ptj}
\end{figure}
\begin{table}[t]
\begin{center}
\begin{tabular}{c c c}
\hline
& 
$\sigma^{\rm{NLO}}$ [fb]& 
$\sigma^{\rm{NLO}}/\sigma^{\rm{LO}}$\\
\hline
$W^-\gamma j$  & $558.7 \pm 2.4$   &  $1.413$\\ 
$W^+\gamma j$ & $676.9 \pm 3.2$   &  $1.339$ \\
\hline
\end{tabular}
\caption{Next-to-leading order cross sections and $K$-factors for the processes $pp\rightarrow W^\pm\gamma j+X$ at the LHC for identified renormalization and factorization scales, $\mu_R=\mu_F=100~\rm{GeV}$. The cuts are chosen as described in the text.}
\label{tab:kfactors}
\end{center}
\end{table}

At leading order, we find a QCD scale dependence of approximately $11\%$ for $W^\pm \gamma j$ production at the LHC, when varying $\mu_R=\mu_F$ by a factor two around $100~\rm{GeV}$, {\it cf.} figs.~\ref{fig:crossecs} and \ref{fig:crossecswp} for identified renormalization and factorization scales. This scale dependence is only reduced to about $7\%$ when including NLO-QCD precision for $W^\pm\gamma j$. This is due to the renormalization scale dependence of the dijet contribution at NLO. Vetoing additional jets results in a stabilization of the cross section, as the veto projects on true $W^\pm \gamma j$ events. This agrees with the results on $W^+W^-j$ production \cite{Dittmaier:2007th}  and $W^\pm\gamma$ production \cite{Baur:1993ir,Dixon:1999di}.

The difference of $W^+\gamma j$ compared to $W^-\gamma j$ is predominantly due to the different parton distribution functions of the dominant subprocesses. Qualitatively, the findings of the $W^-\gamma j$ channel generalize to $W^+ \gamma j$, accompanied by an overall increase of the cross section of about 54\% (see also Tab.~\ref{tab:kfactors}).

At the Tevatron, fig~\ref{fig:crossecstev}, we find a LO scale dependence of $23\%$, which is reduced to about $8\%$ at NLO-QCD. A jet veto is not necessary to stabilize the perturbative corrections as additional jet radiation is sufficiently suppressed by the hard cut on the jet transverse momentum, $p_{T}^{\rm jet}\geq 50~\rm{GeV}$.

For the scale choice $\mu=100~{\rm GeV}$ the total NLO result differs by about $41\%$ for $W^-\gamma j$ from the total LO cross section. As usual, however, the total $K$-factor, defined to be $K=\sigma^{\rm{NLO}}/\sigma^{\rm{LO}}$, reflects only partly the impact of the QCD-quantum corrections on the entire processes' characteristics. Quantitative understanding thereof can be gained from differential $K$-factors of (IR-safe) observables $\cal{O}$,
\be
K({\cal{O}})=\ta{\sigma^{\rm{NLO}}}{\cal{O}}\bigg/\ta{\sigma^{\rm{LO}}}{\cal{O}}.
\ee
In figs.~\ref{fig:ral} and \ref{fig:ptj}, we representatively show the lepton-photon separation and the $p_T$-spectrum of the hardest jet at LO and NLO, accompanied by the respective differential $K$-factors.
The distributions develop significant changes when including NLO-QCD precision, yielding large relative modifications around the total $K$ factor. 
\section{Summary and Outlook}
We have presented first results on the NLO-QCD corrections to $pp\rightarrow W^\pm\gamma j+X$ and $p\bar p\rightarrow W^\pm\gamma j+X$, including
leptonic decays and full off-shell effects for the $W$ boson. 
The calculation has been implemented in a parton-level Monte Carlo program 
based on the \textsc{Vbfnlo} framework which, thus, is fully flexible 
except for the limitation that the Frixione definition of photon isolation 
as given in Eq.~\gl{photonisolation} must be used. Using this program, we 
give sample results for total next-to-leading order cross sections, as 
well as differential distributions and differential $K$-factors.

We find a fairly reduced scale dependence of the total cross sections, {\it cf.} figs.~\ref{fig:crossecs}-\ref{fig:crossecstev}, for a fixed scale choice $\mu_F=\mu_R = \mu$ and our cuts. The corrections turn out to be sizable, around $41\%$ for $W^-\gamma j$ production and $34\%$ for $W^+\gamma j$ production at the LHC. The total correction at the Tevatron is about 30\%.

These total corrections are accompanied by significant modifications of up to $60\%$ for differential distributions when going from LO to NLO, figs.~\ref{fig:ral} and \ref{fig:ptj}.

A more detailed investigation, including analysis of the impact of anomalous couplings and the calculation of NLO-QCD jet veto efficiencies for searches suggested in \cite{Baur:1993ir}, is underway. Eventually, this process will be made publicly available as part of the \textsc{Vbfnlo} package.

\acknowledgements
\noindent We would like to thank Stefan Dittmaier, Stefan Kallweit and Giuseppe Bozzi for helpful discussions, and Steffen Schumann for \textsc{Sherpa}-support.
F.C. acknowledges 
a postdoctoral fellowship of the Generalitat Valenciana ``Beca Postdoctoral d'Excel.l\`encia'' 
and C.~E. is supported by ``\mbox{KCETA} Strukturiertes Promotionskolleg". 
This research is partly funded by the Deutsche Forschungsgemeinschaft 
under SFB TR-9 ``Computergest\"utzte Theoretische Teilchenphysik'', European FEDER and
Spanish MICINN under grant FPA2008-02878, and the Helmholtz alliance ``Physics at the Terascale''.


\end{document}